\documentclass[fleqn,12pt,a4paper]{article} 
\usepackage{amssymb}
\title{\Huge K-Theory Torsion}       
\author{Volker Braun%
  \footnote{volker.braun@physik.hu-berlin.de}} 
\date{\small
  \textit{Humboldt Universit\"at zu Berlin, Institut f\"ur
    Physik\\ Invalidenstra\ss{}e 110, D-10115 Berlin, Germany}}

\newcommand{\dslash}{\mbox{$\partial$ \kern-.92em  \big /}}
\newcommand{\transverse}{\ensuremath{\,\mbox{\lower 0.02ex\hbox{\small$\cap$} 
    \kern-1.03em $\top$}\,}}

\newcommand{\img}{\mathop{\rm img}\nolimits}
\renewcommand{\ker}{\mathop{\rm ker}\nolimits}

\newcommand{\eqdef}{\stackrel{\rm def}{=}}

\newcommand{\R}{\ensuremath{\mathbb R}}
\newcommand{\Q}{\ensuremath{\mathbb Q}}
\newcommand{\Z}{\ensuremath{\mathbb Z}}
\newcommand{\C}{\ensuremath{\mathbb C}}
\newcommand{\N}{\ensuremath{\mathbb N}}
\newcommand{\CP}[1]{\ensuremath{{\C}{\rm P}^{#1}}}
\newcommand{\RP}[1]{\ensuremath{{\R}{\rm P}^{#1}}}
\newcommand{\PGl}{\mathop{\rm PGL}}

\newcommand{\Heven}{\ensuremath{H^\mathrm{ev}}}
\newcommand{\Hodd}{\ensuremath{H^\mathrm{odd}}}
\newcommand{\Ktheory}{K-theory}
\newcommand{\Tor}{\mathop{\rm Tor}\nolimits}
\newcommand{\ch}{\mathop{\rm ch}\nolimits}
\newcommand{\rk}{\mathop{\rm rk}\nolimits}
\newcommand{\spinc}{\ensuremath{\mathrm{Spin_c}}}

\newcommand{\cy}{Calabi-Yau}
\newcommand{\cym}{Calabi-Yau manifold}

\usepackage{theorem}
\theorembodyfont{\rmfamily}

\newtheorem{corollary}{Corollary}
\newtheorem{proof}{Proof}

\begin{document}               
\maketitle                     

\vskip -7.0cm
 \hfill HU-EP-00/24\\
\vskip 7.0cm

\begin{abstract}
  The Chern isomorphism determines the free part of the K-groups from
  ordinary cohomology. Thus to really understand the implications of
  \Ktheory{} for physics one must look at manifolds with K-torsion.
  Unfortunately there are not many explicit examples, and usually for
  very symmetric spaces.  Cartesian products of $\RP{n}$ are examples
  where the order of the torsion part differs between \Ktheory{} and
  ordinary cohomology. The dimension of corresponding branes is also
  discussed. An example for a \cym{} with K-torsion is given.
\end{abstract}

\newpage

\section{Introduction}

It has been shown~\cite{w1}~that \Ktheory{} classifies the topological
charge of the D-brane gauge bundle (or the associated vector bundle).
The crucial observation for this was that adding a brane-antibrane
pair with the same gauge bundle does not change the total charge. Or
in other words, you may add a D-p brane with the trivial bundle, then
try to straighten out any ``windings'', in the bigger bundle, bring
the bundle back in the form where the trivial bundle is one summand,
and then remove the brane you added.

Mathematically, this is called stabilization, and the charge of the
gauge bundle are stable isomorphism classes. It is not difficult to
find examples of vector bundles that are stable isomorphic but not
isomorphic, for example $TS^2$ and the trivial bundle $S^2 \times
\R^2$ (as real vector bundles).

Another way to look at \Ktheory{} is that it is a generalized cohomology
theory, that is it satisfies all the usual axioms except that higher
cohomology groups of a point may not vanish. Since we can express
usual field theory in terms of differential forms and de Rahm
cohomology, it seems natural that a generalization of field
theory leads to a generalized cohomology theory.

Now for all well-behaved spaces $X$ (such as topological manifolds or
finite CW complexes), $K(X)$ is a finitely generated abelian group,
i.e. of the form $\Z^n\oplus\textrm{Torsion}$. Interestingly, torsion
charges can appear. In ordinary field theory you could also have
torsion in integral cohomology $H^\ast(X,\Z)$, but physical fields
must be represented by differential forms, and this prohibits torsion.
But on the \Ktheory{} side torsion charges are apparently physical
charges. The purpose of this paper is to better understand the
relation between integral cohomology and \Ktheory.

{\bf \boldmath For concreteness, I will restrict myself in the
  following to IIB string theory with spacetime manifold $X$, where
  the possible D-brane charges are $K(X)$, the Grothendiek group of
  complex vector bundles.  }

Of course we want $X$ to be a $10$--manifold (where Poincar\'e duality
holds). In the following I will investigate compact topological
manifolds of lower dimension which exhibit torsion in cohomology. The
physical motivation for this is a spacetime of the form $M_d \times
\R^{1,9-d}$, which on the K-theory side is just the $(10-d)$-th suspension
of $M_d$. So the torsion of \Ktheory{} comes purely from the compact
dimensions, and not from the Minkowski part of spacetime.

During the preparation of this paper another work appeared that also
discusses the implications of the Atiyah--Hirzebruch spectral
sequence~\cite{dmw}.

I would like to thank Philip Candelas, Thomas Friedrich, Albrecht
Klemm, Dieter L\"ust, Andr\'e Miemiec, and Ulrike Tillmann.

Here is a quick outline of the following sections:
\begin{enumerate}
\item This introduction
\item The Chern isomorphism and the Atiyah--Hirzebruch spectral
  sequence, which are the main tools used in this paper, are
  introduced. To demonstrate their utility I prove that the order of
  the torsion part is the same in integer cohomology and \Ktheory{}
  for odd $\RP{n}$. The sequence provides a necessary criterion for
  \Ktheory{} torsion.
\item Given some \Ktheory{} element, I determine the dimensionality of
  the corresponding D-brane (That is the minimum dimension needed to
  carry the charge). This can also be calculated from the
  Atiyah--Hirzebruch spectral sequence. 
\item An example with different order of the torsion part in integer
  cohomology an \Ktheory{} is analyzed in detail.
\item By considering line bundles I find a sufficient criterion for
  \Ktheory{} torsion. This makes it possible to give an example for a
  \cym{} with torsion.
\item Conclusion
\end{enumerate}

\section{From $H^\ast(X,\Z)$ to $K^\ast(X)$}

The most important result is the Chern isomorphism:
\begin{eqnarray}
  \label{chern}
  K^0(X) \otimes_\Z \R &\simeq& \bigoplus H^{2i}(X,\R)
  \quad \eqdef \quad \Heven(X,\R)
  \\ \nonumber 
  K^1(X) \otimes_\Z \R &\simeq& \bigoplus H^{2i+1}(X,\R)  
  \quad \eqdef \quad \Hodd(X,\R)
\end{eqnarray}
which is induced by the Chern character $\ch:K(X)\to \Heven(X,\R)$. 

This means that we can compute the free part of \Ktheory{} directly
from ordinary de Rahm cohomology. Or in physical language, \Ktheory{}
without torsion is just a reformulation of what one already knows from
calculations on the level of differential forms. On the other hand
side the torsion part of $\Heven(X,\Z)$ and $K(X)$ do in general
differ, for example
\footnote{$\RP{5}$ is not spin, and therefore not a phenomenologically
  viable background spacetime. $\RP{7}$ would be a counterexample that
  is spin. 

  Remember the relevant Stiefel-Whitney classes
  $w_1(\RP{n})=n+1$, $w_2(\RP{n})=\frac{n(n+1)}{2}$ (mod 2) }
\begin{eqnarray}
  \label{eq:rp5}
  K(\RP{5}) &=& \Z \oplus \Z_4
  \\ \nonumber 
  \Heven(\RP{5},\Z) &=& \Z \oplus \Z_2 \oplus \Z_2 
\end{eqnarray}
It has been noted~\cite{hk}~that --- although there is no surjective
group homomorphism --- the order of the torsion part is
equal. Unfortunately, this is caused by peculiarities in the
cohomology of real projective spaces and not a generic
feature. A counterexample will be presented in section
\ref{sec:examples}.

For now, lets use the Atiyah--Hirzebruch spectral
sequence~\cite{sequence}~to understand why the order is indeed equal
for $\RP{n}$, with $n$ odd so that the manifold is orientable. This
spectral sequence stems from the filtration of the space by its
CW-skeleton. It has the second term
\begin{equation}
  \label{eq:E2}
  E_2^{p,q}= \left\{
    \begin{array}{cl}
      H^p(\RP{n},\Z) & q~\textrm{even} \\
      0 & q~\textrm{odd} \\
    \end{array} 
    \right.
\end{equation}
and converges towards the associated graded complex of $K(\RP{n})$. 

For simplicity, take $n=5$:
\begin{equation}
  E_2^{p,q} = 
  \begin{array}{c|cccccc}
    \uparrow &  0 & 0 &    0 & 0 &    0 &  0 \\
    q & \Z & 0 & \Z_2 & 0 & \Z_2 & \Z \\
    &  0 & 0 &    0 & 0 &    0 &  0 \\
    & \Z & 0 & \Z_2 & 0 & \Z_2 & \Z \\
    &  0 & 0 &    0 & 0 &    0 &  0 \\
    & \Z & 0 & \Z_2 & 0 & \Z_2 & \Z \\
    \cline{2-7} 
    \multicolumn{7}{r}{p \rightarrow}
  \end{array}
\end{equation}
The differential $d_2^{p,q}:E_2^{p,q}\to E_2^{p+2,q-1}$ either has
domain or range $0$, thus 
\begin{equation}
  E_3^{p,q} = \ker d_2^{p,q}/\img d_2^{p-2,q+1} = E_2^{p,q} 
\end{equation}
In this sequence the $d_\mathrm{even}$ are obviously
irrelevant, and $E_{2k} = E_{2k+1}$.

The only\footnote{Of course the table is cyclic of order 2 in $q$}
$d_3^{p,q}:E_3^{p,q}\to E_3^{p+3,q-2}$ with nonvanishing domain and
range is\footnote{In~\cite{sequence}~it is noted without proof that
  $d_3=\mathop{Sq^3}$, the third Steenrod Square} $d_3^{2,2}:\Z_2\to
\Z$. Since there is no nonzero group homomorphism from $\Z_2$ to $\Z$,
$d_3=0$.

So far we found $E_5=E_2$, and again there is only one $d_5$ with
nonvanishing domain and range, $d_5^{0,4}:\Z\to \Z$. But the Chern
isomorphism tells us that after tensoring everything with $\Q$ the
spectral sequence already degenerates at level 2. Thus $d_5 \otimes_\Z
\Q = 0$, and since domain is torsion free this implies $d_5=0$ (This
also proves that torsion in cohomology is necessary for torsion in
\Ktheory).

Thus $E_\infty = E_2$, but this is not enough to compute $K(\RP{5})$.
All that it tells us is that there is a filtration
\begin{eqnarray}
  \label{eq:Kfilt}
  K^0(\RP{5}) &=& F^0_6 \supset F^0_5 \supset F^0_4 
  \supset F^0_3 \supset F^0_2 \supset F^0_1 \supset 0 
  \\ \nonumber
  K^1(\RP{5}) &=& F^1_6 \supset F^1_5  \supset F^1_4 
  \supset F^1_3 \supset F^1_2 \supset F^1_1 \supset 0 
\end{eqnarray}
such that the successive cosets are the even respectively odd
diagonals of $E_\infty$:
\begin{equation}
  \label{eq:Kfilt_ex}
  \begin{array}{cccccc}
    F^0_6/F^0_5 = \Z & 
    F^0_5/F^0_4 = 0 & 
    F^0_4/F^0_3 = \Z_2 & 
    F^0_3/F^0_2 = 0 & 
    F^0_2/F^0_1 = \Z_2 & 
    F^0_1/0 = 0 \\
    F^1_6/F^1_5 = 0 & 
    F^1_5/F^1_4 = 0 & 
    F^1_4/F^1_3 = 0 & 
    F^1_3/F^1_2 = 0 & 
    F^1_2/F^1_1 = 0 & 
    F^1_1/0 = \Z 
  \end{array}  
\end{equation}
Obviously $K^1(\RP{5}) = F^1_6 = F^1_5 = F^1_4 = F^1_3 = F^1_2 = F^1_1
= \Z$. But for $K^0(\RP{5})$ we find $F^0_1=0$, $F^0_3=F^0_2=\Z_2$ and
then hit the extension problem: either $\Z_4/\Z_2 = \Z_2$ or $(\Z_2
\oplus \Z_2)/\Z_2 = \Z_2$. So $F^0_5 = F^0_4 =
\Z_4~\textrm{or}~\Z_2\oplus\Z_2$. Since $F^0_5$ is pure torsion each
possibility determines a unique $K(\RP{5})=F^0_6$, either
$\Z\oplus\Z_4$ or $\Z\oplus\Z_2\oplus\Z_2$.

However, the ambiguity is between groups of equal order (since the
ambiguous extension was between finite abelian groups), and moreover
one of the possibilities was $\Heven(\RP{5})$, since the spectral
sequence already degenerates at $E_2$. 

The same argument can be used for all real projective spaces to prove
that the order of the torsion part of cohomology and \Ktheory{} are
equal, but as we have seen the proof depends on the special properties
of \RP{n}.

\section{Dimension of D-branes}

\subsection{Filtering $K(X)$}

In flat space one can explicitly construct vector bundles that carry a
nontrivial topological charge (using the Clifford algebra,
see~\cite{w1}~and~\cite{abs}).  The bundle is trivial everywhere
except on a hyperplane of even codimension, which is identified with
the D-brane.  One can extend this construction to general submanifolds
with \spinc{} normal bundle. This fits nicely to the fact that
D-branes in IIB are even dimensional.

But to understand what the charges are one should rather understand
which submanifold can carry a given \Ktheory{} element.  The intuitive
answer would be: An arbitrary submanifold $Y\subset X$ can carry the
charge $x = [E]-[F] \in K(X)$ if there exists an isomorphism $E|_{X-Y}
\simeq F|_{X-Y}$.  Of course this is not well-defined, since the same
\Ktheory{} element could be represented by different vector bundles
$E'$, $F'$ that are stably isomorphic but not isomorphic. So we should
really ask whether there exists an isomorphism $(E|_{X-Y}\oplus \C^k)
\simeq (F|_{X-Y}\oplus \C^k)$ for some $k\in \Z$.

One would like to use the inclusion map $i:X-Y\hookrightarrow X$ to pull
back $x$, and thus automatically include stabilization as an element
of $K(X-Y)$, but unfortunately in general $i^\ast(E)-i^\ast(F) \not\in
K(X-Y)$ since the complement $X-Y$ is not compact (Remember that
\Ktheory{} on noncompact spaces are differences of vector bundles that
are isomorphic outside a compact subset). 

So instead take compact submanifolds $j:Z\hookrightarrow X$ as probes:
If their dimension is too low, they will generically miss the D-brane
and the pullback $j^\ast(x)=0\in K(Z)$. Since $j^\ast$ depends only on
the homotopy class of $j$, we do not have to worry about degenerate
cases. If we cannot detect $x$ with submanifolds of a given dimension
$p$ then we conclude that $x$ is carried by a D-brane of codimension
greater than $p$.

But the total charge $0\in K(X)$ could also be carried by a
brane-antibrane pair that is separated in spacetime. Probing only in
the neighborhood of one brane one would falsely find a charge. So our
probe submanifold must somehow be big enough. Discussing this in terms
of submanifolds is very cumbersome, so instead think of spacetime $X$
as a cell complex (simplicial complex or CW complex). Then take the
$p$-skeleton $X^p$ as probe; it can easily be seen that this is
independent of the chosen cell structure. Any cell complex embedded in
$X$ is a subcomplex for some cell structure on $X$, in that sense
$X^p$ probes the whole space.

Let $K_p(X)$ be the subgroup of $K(X)$ of charges that live on a
brane\footnote{More precisely a stack of coincident branes, although
  I will not make that distinction in the following} of codimension
$p$ or higher, that is D-$(\dim(X)-p-1)$-branes or lower. According
to the previous arguments
\begin{equation}
  \label{eq:Kp}
  K_p(X) = \ker \Big( K(X) \to K(X^{p-1}) \Big)
\end{equation}
where the map is the one induced by the inclusion $X^{p-1}
\hookrightarrow X$.

This yields a filtration
\begin{equation}
  K(X)=K_{0}(X) ~\supset~ 
  \tilde{K}(X)=K_1(X) ~\supset~ 
  K_2(X)  ~\supset \cdots \supset~ 
  K_{\dim X+1}(X)=0
\end{equation}
where the successive quotients $K_p(X)/K_{p+1}(X)$ are the
D-$(\dim(X)-p-1)$-brane charges.

\subsection{Remarks}

Lets try to understand eq.~(\ref{eq:Kp}) better. $K_{\dim X+1}(X)=0$ means
that there are no D-$(-2)$-branes or less, which is correct. 

The $9$-brane charges ($p=0$) are $K(X)/\tilde{K}(X)=\Z$, which is the
virtual rank of the bundle pair. This we can also understand: If we do
not start with the same number of $9$ and $\bar{9}$-branes, then there
will always be a $10$-dimensional brane left. On the other hand side
if the virtual rank is $0$ (as required by tadpole cancellation), then
the vector bundles are isomorphic over sufficiently small open sets
(since they are locally trivial), which one could use to localize the
nontrivial windings at a subspace of codimension $1$.

Fortunately there is a way to calculate the quotients
$K_p(X)/K_{p+1}(X)$. First note that one can extend eq.~(\ref{eq:Kp})
to the higher $K$-groups straightforwardly:
\begin{equation}
  K^n_p(X) = \ker \Big( K^n(X) \to K^n(X^{p-1}) \Big)  
\end{equation}
And the associated graded complex to this filtration is precisely the
limit of the Atiyah--Hirzebruch spectral sequence
\begin{equation}
  \label{eq:Einfty}
  E_\infty^{p,q} = K^{p+q}_{p}(X)/K^{p+q}_{p+1}(X)
\end{equation}

If there is no torsion in integer cohomology then the spectral
sequence degenerates at level 2, and (compare eq.~(\ref{eq:E2}))
\begin{equation}
  K^\ast_{p}(X)/K^\ast_{p+1}(X) \simeq H^p(X,\Z)
\end{equation}
where the isomorphism is just the Chern character. This confirms the
interpretation of the dimensionality of the \Ktheory{} elements.

The odd rows in $E_\infty^{p,q}$ all vanish, so for odd $p$ and odd
$q$
\begin{equation}
  K^{p+q}_{p}(X)/K^{p+q}_{p+1}(X) = K_p(X)/K_{p+1}(X) = 0
\end{equation}
which just means that there are no topological charges for odd
dimensional D-branes. 

A word of caution: even if $K(X)$ is torsion-free, one of the
successive quotients can be torsion, as in the example $\Z/2\Z =
\Z_2$. Physically, this means that there can be an apparent torsion
charge on a D-brane in the sense that multiple copies of that brane
can decay to something lower-dimensional, which a single brane cannot.
But the lower-dimensional remnant then carries an ordinary
(non-torsion) charge that keeps track of the number of branes we
started with.

\section{Examples}
\label{sec:examples}

\subsection{$K(\RP{n})$}

The best way to construct manifolds with K-torsion is to use quotients
of well-understood manifolds (like the sphere) by free group
actions. At the example \RP{n} I will review the necessary tools (See
e.g.~\cite{k}).

Let $\Z_2$ act on $S^n$ ($n$ odd) via the antipodal map, a free group
action. In general (for free group actions) \Ktheory{} on the quotient
is equal to equivariant \Ktheory{} on the covering space
$K^\ast(S^n/\Z_2)=K_{\Z_2}^\ast(S^n)$. Writing down the (cyclic) long
exact sequence associated to the inclusion $S^n \hookrightarrow
D^{n+1}$, we find:
\begin{equation}
  \label{eq:RP3_1}
  \begin{array}{ccccc}
    K^1_{\Z_2}(S^n) & \longleftarrow &
    K^1_{\Z_2}(D^{n+1}) & \longleftarrow &
    K^1_{\Z_2}(D^{n+1},S^n)
    \\
    \downarrow & & & & \uparrow  \\
    K^0_{\Z_2}(D^{n+1},S^n) & \longrightarrow &
    K^0_{\Z_2}(D^{n+1}) & \longrightarrow &     
    K^0_{\Z_2}(S^n)
  \end{array}
\end{equation}
where the $\Z_2$ action on the disk $D^{n+1}$ is the obvious extension
of the $\Z_2$--action on $S^n$.

Now identify $K^0_{\Z_2}(D^{n+1},S^n)$, virtual differences of
vector bundles on $D^{n+1}$ that are isomorphic over the boundary,
with $K^0_{\Z_2}(\R^{n+1})$, virtual differences on $\R^{n+1}$ with
isomorphism outside a compact subset. The associated $\Z_2$--action on
$\R^{n+1}$ is again $x\mapsto -x$. Since $n+1$ is even, we can interpret
$\R^{n+1}=\C^{(n+1)/2}\eqdef \C^m$ with a linear $\Z_2$--action on
$\C^m$. And this is a $\Z_2$--equivariant vector bundle over a point.

Then use the Thom isomorphism, that is $K_G(E)=K_G(X)$ for any
$G$--vector bundle $E$ over $X$ (as abelian groups, the multiplication
law is different):
\begin{equation}
  K^0_{\Z_2}(D^{n+1},S^n) = 
  K^0_{\Z_2}(\C^m) =
  K^0_{\Z_2}(\{\textrm{pt}\}) = 
  R(\Z_2) = \Z[x]/x^2-1
\end{equation}
$R(\Z_2)$ are the formal differences of representations of $\Z_2$
(with the obvious ring structure induced by the tensor product of
representations), and $x$ denotes the unique nontrivial irreducible
representation of $\Z_2$. If one is only interested in the underlying
abelian group, this is of course $\Z\oplus\Z$. 

Doing the same for $K^1$ and using the homotopy $D^{n+1}\sim \{pt\}$,
we evaluate eq.~(\ref{eq:RP3_1}):
\begin{equation}
  \label{eq:RP3_2}
  \begin{array}{ccccc}
    K^1_{\Z_2}(S^n) & \longleftarrow &
    0 & \longleftarrow &
    0
    \\
    \downarrow & & & & \uparrow  \\
    \Z[x]/x^2-1 & \stackrel{f}{\longrightarrow} &
    \Z[x]/x^2-1 & \stackrel{g}{\longrightarrow} &     
    K^0_{\Z_2}(S^n)
  \end{array}
\end{equation}
Since $\Z[x]/x^2-1$ is torsion free as abelian group, so must be
$K^1_{\Z_2}(S^n)=K^1(\RP{n})$. From the Chern isomorphism then follows
that $K^1(\RP{n})=\Z$. But to determine the torsion part of
$K^0(\RP{n})$, we need to identify the map $f$. Tracing everything
back to the Thom isomorphism, one can show that $f$ is multiplication
with $(x-1)^m$. Using exactness ($\img f = \ker g$) we find
\begin{eqnarray}
  K^0(\RP{n}) &=& K^0_{\Z_2}(S^n) =
  \Z[x] / \left< x^2-1, (x-1)^m \right> =
  \nonumber \\ &=&
  \Z[z] / \left< (z+1)^2-1, z^m \right> =
  \Z[z] / \left< z^2+2z, z^m \right>
\end{eqnarray}
Up to the given relations, each ring element can be represented as
$az+b$, $a,b\in \Z$. While $b$ is not subject to any relation, we can
use $z^2+2z=0$ and $z^m=0$ to show $2^{m-1}z=0$. Therefore (ignoring
the ring structure):
\begin{eqnarray}
  K^1(\RP{n}) &=& \Z 
  \nonumber \\
  K^0(\RP{n}) &=& \Z \oplus \Z_{2^{m-1}}
\end{eqnarray}

\subsection{$K(\RP{3}\times \RP{5})$}

Here is the promised example of a space where the order of the torsion
subgroup in \Ktheory{} and ordinary cohomology differs.

Of course we use the K\"unneth formula to calculate the
cohomology\footnote{In this section, $H^\ast(X)$ is always cohomology
  with integer coefficients} of a Cartesian product:
\begin{equation}
  \label{eq:kuenneth_cohomology}
  \hspace{-1cm}
  0 \longrightarrow
  \bigoplus_{i+j=m} H^i(X) \otimes H^j(Y) \longrightarrow
  H^m(X\times Y) \longrightarrow
  \bigoplus_{i+j=m+1} \Tor\left(H^i(X),H^j(Y)\right)
  \longrightarrow 0
\end{equation}
The cohomology of real projective space is
\begin{equation}
  H^i(\RP{3})=\left\{
    \begin{array}{cl}
      \Z   & i=3 \\
      \Z_2 & i=2 \\
      0    & i=1 \\
      \Z   & i=0 \\
    \end{array} \right. 
  \qquad
  H^i(\RP{5})=\left\{
    \begin{array}{cl}
      \Z   & i=5 \\
      \Z_2 & i=4 \\
      0    & i=3 \\
      \Z_2 & i=2 \\
      0    & i=1 \\
      \Z   & i=0 \\
    \end{array} \right. 
\end{equation}
Thus eq.~(\ref{eq:kuenneth_cohomology}) contains the exact sequences
\begin{equation}
  \begin{array}{ccccccccc}
    0 
    &\longrightarrow& 
    \Z
    &\longrightarrow& 
    H^8(\RP{3}\times \RP{5})
    &\longrightarrow&
    0
    &\longrightarrow&
    0 \smallskip\\
    0 
    &\longrightarrow& 
    \Z_2 \oplus \Z_2
    &\longrightarrow& 
    H^7(\RP{3}\times \RP{5})
    &\longrightarrow&
    0
    &\longrightarrow&
    0 \smallskip\\
    0 
    &\longrightarrow& 
    \Z_2
    &\longrightarrow& 
    H^6(\RP{3}\times \RP{5})
    &\longrightarrow&
    0
    &\longrightarrow&
    0 \smallskip\\
    0 
    &\longrightarrow& 
    \Z \oplus \Z_2
    &\longrightarrow& 
    H^5(\RP{3}\times \RP{5})
    &\longrightarrow&
    \Z_2
    &\longrightarrow&
    0 \smallskip\\
    0 
    &\longrightarrow& 
    \Z_2 \oplus \Z_2
    &\longrightarrow& 
    H^4(\RP{3}\times \RP{5})
    &\longrightarrow&
    0
    &\longrightarrow&
    0 \smallskip\\
    0 
    &\longrightarrow& 
    \Z
    &\longrightarrow& 
    H^3(\RP{3}\times \RP{5})
    &\longrightarrow&
    \Z_2
    &\longrightarrow&
    0 \smallskip\\
    0 
    &\longrightarrow& 
    \Z_2 \oplus \Z_2
    &\longrightarrow& 
    H^2(\RP{3}\times \RP{5})
    &\longrightarrow&
    0
    &\longrightarrow&
    0 \smallskip\\
    0 
    &\longrightarrow& 
    0
    &\longrightarrow& 
    H^1(\RP{3}\times \RP{5})
    &\longrightarrow&
    0
    &\longrightarrow&
    0 \smallskip\\
    0 
    &\longrightarrow& 
    \Z
    &\longrightarrow& 
    H^0(\RP{3}\times \RP{5})
    &\longrightarrow&
    0
    &\longrightarrow&
    0 
  \end{array}
\end{equation}
Using Poincar\'e duality ($\RP{3}\times \RP{5}$ is an orientable
8-manifold since each factor is), $H^5_\mathrm{tors}\simeq
H^4_\mathrm{tors}$ and $H^3_\mathrm{tors}\simeq H^6_\mathrm{tors}$.
This fixes the extension ambiguities, and we find
\begin{equation}
  H^i(\RP{3} \times \RP{5})=\left\{
    \begin{array}{cl}
      \Z                          & i=8 \\
      \Z \oplus \Z_2 \oplus \Z_2  & i=7 \\
      \Z \oplus \Z_2              & i=6 \\
      \Z \oplus \Z_2 \oplus \Z_2  & i=5 \\
      \Z_2 \oplus \Z_2            & i=4 \\
      \Z \oplus \Z_2              & i=3 \\
      \Z_2 \oplus \Z_2            & i=2 \\
      0                           & i=1 \\
      \Z                          & i=0 \\
    \end{array} \right. 
  \quad \Rightarrow~
  \left\{
    \begin{array}{c}
      \Heven(\RP{3}\times \RP{5}) = 
      \Z^2 \oplus \Z_2^5
      \smallskip\\
      \Hodd(\RP{3}\times \RP{5}) = 
      \Z^2 \oplus \Z_2^5
    \end{array}
  \right.
\end{equation}
For \Ktheory{} there is the following~\cite{kunneth}~analog to the
ordinary K\"unneth formula:
\begin{equation}
  \label{eq:K_kuenneth}
  \hspace{-1cm}
  0 \longrightarrow
  \bigoplus_{i+j=m} K^i(X) \otimes K^j(Y) \longrightarrow
  K^m(X\times Y) \longrightarrow
  \bigoplus_{i+j=m+1} \Tor\left(K^i(X),K^j(Y)\right)
  \longrightarrow 0
\end{equation}
where all indices are modulo $2$. Thus
\begin{equation}
  \hspace{-1cm}
  \begin{array}{ccccccccc}
    0 
    &\longrightarrow& 
    \Big[\Z \otimes \big(\Z\oplus \Z_4\big)\Big] \oplus
    \Big[\big(\Z\oplus \Z_2\big) \otimes \Z\Big]
    &\longrightarrow& 
    K^1(\RP{3}\times \RP{5})
    &\longrightarrow&
    \Z_2
    &\longrightarrow&
    0 \\
    & & \shortparallel \\
    & & \Z \oplus \Z \oplus \Z_4 \oplus \Z_2 
    \medskip  \\ 
    0 
    &\longrightarrow& 
    \Big[\Z \otimes \Z\Big] \oplus
    \Big[\big(\Z\oplus \Z_2\big) 
    \otimes \big(\Z \oplus \Z_4\big)\Big]
    &\longrightarrow& 
    K^0(\RP{3}\times \RP{5})
    &\longrightarrow&
    0
    &\longrightarrow&
    0 \\
    & & \shortparallel \\
    & & \Z \oplus \Z \oplus \Z_4 \oplus \Z_2 \oplus \Z_2
  \end{array}    
\end{equation}
Using the duality~\cite{wm}\footnote{I am grateful to Ulrike Tillmann for
  sketching to me how one could give a rigorous proof} between the
torsion part of $K^0$ and $K^1$ for an even-dimensional orientable
manifold, we arrive at the following result:
\begin{eqnarray}
  K^1(\RP{3}\times \RP{5}) &=& \Z^2 \oplus \Z_4 \oplus \Z_2^2 
  \nonumber \\
  K^0(\RP{3}\times \RP{5}) &=& \Z^2 \oplus \Z_4 \oplus \Z_2^2 
\end{eqnarray}
The order of the torsion subgroups of $K^0$ and \Heven{} does not
match. Tracing it back through our calculation, we see that this stems
from the well-known fact that the order of the torsion of a tensor
product is not determined by the orders of the torsion subgroups of
the factors. To be precise $\Z_2 \otimes \Z_4 = \Z_2$, while $\Z_2
\otimes (\Z_2 \oplus \Z_2)=\Z_2 \oplus \Z_2$.

\subsection{Complete Intersections}
\label{sect:ci}

For physical reasons it would be nice if the underlying space is
\cy. Unfortunately hypersurfaces in toric varieties have
torsion-free \Ktheory: 

The smooth toric variety (of complex dimension $m$) does not have
torsion in integer homology. The Lefschetz hyperplane theorem yields
torsion free homology of the hypersurface in (real) dimensions $0$ to
$m-1$. But Poincar\'e duality then fixes the torsion part of the
whole homology, since $H_i(X,\Z)_\mathrm{tors}\simeq H_{\dim
  X-1-i}(X,\Z)_\mathrm{tors}$.  Duality with integer cohomology then
gives rise to torsion free cohomology. But torsion in integer
cohomology is necessary for \Ktheory{} torsion.

\section{Multiply Connected Spaces}

We have seen that integer cohomology provides a necessary although not
sufficient tool to determine whether a given manifold has torsion in
\Ktheory. The purpose of this section is to give an easy sufficient
criterion. The idea is that line bundles are stably isomorphic if and
only if they are isomorphic, so stability is not a relevant concept
for one-dimensional vector bundles. Then we just have to construct
line bundles where a certain finite sum is (stable) trivial. This
happens if the first Chern class $c_1 \in H^2(X,\Z)$ is torsion.

\subsection{Line Bundles}

Let us have a closer look to the aforementioned properties of line
bundles. A $n$-dimensional vector bundle is in general defined via its
transition functions on some open cover $X= \cup_{i \in I}U_i$:
\begin{equation}
  g_{ij}: U_i \cap U_j \to U(n,\C)
\end{equation}
For a line bundle, this means 
\begin{equation}
  g_{ij}: U_i \cap U_j \to U(1)
\end{equation}
Now two line bundles $L_1,L_2$ (with transition functions $g^{(1)},
g^{(2)}$) are stably isomorphic if there exists
an $n\in \N$:
\begin{equation}
  L_1 \oplus \C^n \simeq L_2 \oplus \C^n
\end{equation}
But the determinant bundle of a line bundle plus a trivial bundle is
again the line bundle. Remember that the transition functions
$\tilde{g}_{ij}^{(k)}$ of the determinant bundle $\wedge^{n+1} (L_k \oplus
\C^n)$ are the determinants of the transition function matrices of
$L_k \oplus \C^n$:
\begin{equation}
  \tilde{g}_{ij}^{(k)} = 
  \det \left(
    \begin{array}{cccc}
      g_{ij}^{(k)} & & & \\
       &1 & & \\
       & &\ddots & \\
       & & &1 \\
    \end{array}
    \right)
    = g_{ij}^{(k)}
    \qquad
    k=1,2
\end{equation}
Therefore
\begin{equation}
  L_1 \oplus \C^n \simeq L_2 \oplus \C^n
  \quad \Rightarrow \quad 
  \wedge^{n+1} (L_1 \oplus \C^n) \simeq \wedge^{n+1} (L_2 \oplus \C^n)
  \quad \Rightarrow \quad 
  L_1 \simeq L_2
\end{equation}
Of course the ``$\Leftarrow$'' is trivial.

By a standard argument we identify then the isomorphism classes of
transition functions of line bundles with the Cech cohomology group
$H^1\left(X,C^0(U(1))\right)$, where $C^0(U(1))$ is the sheaf of
$U(1)$-valued continuous functions. The long exact sequence associated
to the exponential short (sheaf) exact sequence is then
\begin{equation}
  \label{eq:les}
  \cdots 
  \rightarrow 
  H^1\big(X,C^0(\R)\big)
  \rightarrow
  H^1\Big(X,C^0\big(U(1)\big)\Big) 
  \stackrel{c_1}{\rightarrow}
  H^2(X,\Z)
  \rightarrow
  H^2\big(X,C^0(\R)\big)
  \rightarrow
  \cdots
\end{equation}
which yields the desired isomorphism since $C^0(\R)$ is a fine sheaf,
$H^i\big(X,C^0(\R)\big)=0 ~\forall i\geq 1$.

\subsection{Adding line bundles}

Now assume $E$ is a line bundle on $X$ with $0\not= c_1(E) \in
H^2(X,\Z)$ pure torsion (according to the previous section then
$[E]-[1]\not= 0 \in K(X)$). But observe that the group law in $K(X)$
is based on the Whitney sum $E\oplus E$, while the group law in
$H^2(X,\Z)$ corresponds to the tensor product\footnote{By the
  isomorphism in eq.~(\ref{eq:les}) this is the group law in
  $H^1(X,C^0(U(1)))$, which corresponds to multiplying the $U(1)$
  transition functions} $E \otimes E$. And of course $[E]\in K(X)$
does not generate a torsion subgroup since
\begin{equation}
  \dim( n [E] ) = 
  \dim\Big(
  \underbrace{ [E] + \cdots + [E] }_{n~\mathrm{times}}
  \Big) = 
  n \not=0 \qquad \forall n \in \Z-\{0\}
\end{equation}
However $[E]-[1]\in K(X)$ is a torsion element ($1$ denotes the
trivial line bundle). This follows from the Chern isomorphism:
\begin{corollary}
  Let $0\not= x \in K(X)$. Then $x$ is a torsion element if and only
  if $\ch(x)=0$.
\end{corollary}
\begin{proof}  
  \begin{itemize}
  \item ``$\Rightarrow$'': Since $\ch:K(X)\to \Heven(X,\R)$ is a group
    homomorphism this is trivial.
  \item ``$\Leftarrow$'': Assume that $x\in K(X)$ is free but
    $\ch(x)=0$. Thus $\dim(\img(\ch))<\rk(K(X))$, in contradiction to
    the Chern isomorphism (eq.~(\ref{chern})).
  \end{itemize}
\end{proof}

In our case the Chern character $\ch(E) = e^{c_1(E)} = 1 + c_1(E) +
\cdots = 1 \in \Heven(X,\R)$ since $c_1(E)$ was assumed to be a
torsion element in $H^2(X,\Z)$ (so its image in $H^2(X,\R)$ vanishes).
Therefore $\ch([E]-[1]) = \ch(E)-\ch(1) = 0$ and $[E]-[1]$ generates a
nontrivial torsion subgroup.

\subsection{Multiply connected \cym{}s: Quintics}

Consider the Fermat quintic $Y\subset\CP{4}$:
\begin{equation}
  \label{eq:quintic}
  \sum_{i=1}^5 z_i^5 = 0
\end{equation}
with the $\Z_5=G=\{1,g,g^2,g^3,g^4\}$ symmetry generated by 
\begin{equation}
  g:~ \left[ z_1 : z_2 : z_3 : z_4 : z_5 \right]
  \mapsto
  \left[ z_1 : \alpha z_2 : \alpha^2 z_3 : 
    \alpha^3 z_4 : \alpha^4 z_5 \right]
  \qquad \alpha = e^\frac{2 \pi i}{5}
\end{equation}
The group $G$ acts freely on $Y$: The only fixed point $[1:0:0:0:0]\in
\CP{4}$ of the ambient space is missed by the hypersurface
eq.~(\ref{eq:quintic}).

This means that the quotient $X=Y/G$ is a (nonsingular) \cym{}. The
quotient is still projective algebraic, but of course not a complete
intersection since this would contradict section~\ref{sect:ci}; this
simply means that it is a hypersurface in some projective space where
one cannot eliminate all equations.

Since the quintic $Y$ was simply connected (as every complete
intersection), we can determine the quotient's fundamental group from
the long exact homotopy sequence (for $Y$ as a bundle over $X$ with
fiber $G$):
\begin{equation}
  \cdots ~\to~
  \underbrace{\pi_1(G)}_{=0} ~\to~ 
  \underbrace{\pi_1(Y)}_{=0} ~\to~ 
  \pi_1(X) ~\to~ 
  \underbrace{\pi_0(G)}_{=G} ~\to~ 
  \underbrace{\pi_0(Y)}_{=0} ~\to~ 
  \underbrace{\pi_0(X)}_{=0}
\end{equation}
Since $Y$ is a complete intersection,
$h^{1,1}(Y)=h^{1,1}(\CP{4})=1$. The quotient $X$ is still K\"ahler
(the K\"ahler class $\omega = \partial \bar{\partial} \log ||Z||^2$ is
$G$-invariant), so that $h^{1,1}(X)=h^{1,1}(Y)=1$.

The complex structure deformations $h^{2,1}(Y)$ correspond to the
monomials modulo $\PGl(4)$ (the automorphisms of the ambient space)
and rescaling of the equation. Here there are ${5+5-1 \choose 5} =
126$ monomials, and $|\PGl(4)|=24$. Therefore $h^{2,1}(Y) = 126 - 24 -
1=101$. The complex structure deformations of the quotient are the
$G$-invariant monomials, straightforward counting gives $26$. But now
by treating every coordinate separately in the $G$-action the full
$\PGl(4)$ is broken to the diagonal subgroup (4 parameters). Therefore
$h^{2,1}(X)=26-4-1=21$. This is confirmed by the Euler number
\begin{equation}
  \chi(Y) = 2\Big(h^{2,1}(Y)-h^{1,1}(Y)\Big) = 200
  \qquad 
  \chi(X) = 2\Big(h^{2,1}(X)-h^{1,1}(X)\Big) = 40
\end{equation}
As we expect for a free $\Z_5$ group action, $\chi(Y) = 5 \chi(X)$.
The Hodge diamond 
\begin{equation}
  h^{p,q}(X) = \quad
  \begin{array}{ccccccc}
    & & &1& & & \\
    & &0& &0& & \\
    &0& &1& &0& \\
   1& &21& &21& &1\\
    &0& &1& &0& \\
    & &0& &0& & \\
    & & &1& & & 
  \end{array}
\end{equation}
determines the free part of integer cohomology, now we have to find
the torsion part. For every manifold $H^1(X,\Z)$ is torsion free,
since the torsion part is dual to the torsion part in $H_0(X,\Z)=\Z$.
Furthermore $H_1(X,\Z)$ is the abelianization of $\pi_1(X)=\Z_5$,
which was already abelian. Therefore $H_1(X,\Z)=\Z_5$. By the
universal coefficient theorem $H^2(X,\Z)_\mathrm{tors} \simeq
H_1(X,\Z)_\mathrm{tors}=\Z_5$.

The hard part is the torsion in $H^3$ (Poincar\'e duality then
determines the rest). We are going to use the following sequence
\cite{eilenbergmclane}:
\begin{equation}
  \label{eq:eilenbergmclane}
  0 ~\to~ \Sigma_2 ~\to~ H_2(X,\Z) ~\to~ H_2(\Z_5) ~\to~ 0
\end{equation}
where\footnote{This is corrected version of the sequence in
  \cite{rings}} $\Sigma_2$ is the image of $\pi_2(X)$ in $H_2(X,\Z)$.
With other words $\Sigma_2$ are the homology classes that can be
represented by $2$--spheres.

So we need to determine $\pi_2(X)$ first. We know that on the covering
space $\pi_2(Y) = H_2(Y) = \Z$ (The Hurewicz isomorphism theorem) since
$Y$ is simply connected. But every map $f:S^2 \to X$ can be lifted to
$\tilde{f}:S^2 \to Y$ since the $S^2$ is simply connected. That is the
$S^2$ cannot wrap the nontrivial $S^1\subset X$. More formally we can
use the homotopy long exact sequence:
\begin{equation}
  \cdots ~\to~
  \underbrace{\pi_2(G)}_{=0} ~\to~ 
  \pi_2(Y) ~\to~ 
  \pi_2(X) ~\to~ 
  \underbrace{\pi_1(G)}_{=0} ~\to~ 
  \cdots
\end{equation}
to show that $\pi_2(X) = \pi_2(Y) = \Z$.

The group homology $H_2(\Z_5)=0$, therefore
eq.~(\ref{eq:eilenbergmclane}) determines an isomorphism $\Sigma_2
\simeq H_2(X,\Z)$. We know already that the free part
$H_2(X,\Z)_\mathrm{free} = \Z$ from the Hodge diamond. But then the
map $\pi_2(X) \to \Sigma_2$ must have been injective since the domain
is $\Z$ and the image at least $\Z$. Therefore $\Sigma_2 = \Z$ and the
torsion part $H^3(X,\Z)_\mathrm{tors} \simeq H_2(X,\Z)_\mathrm{tors} =
0$. 

We have seen that 
\begin{equation}
  \label{eq:quintic_homology}
  H^i(X,\Z) = 
  \left\{
    \begin{array}{cl}
      \Z               & i=6 \\
      \Z_5             & i=5 \\
      \Z               & i=4 \\
      \Z^{44}          & i=3 \\
      \Z\oplus \Z_5    & i=2 \\
      0                & i=1 \\
      \Z               & i=0 \\
    \end{array}
  \right.
\end{equation}
From the Atiyah--Hirzebruch spectral sequence it is obvious that
either the $\Z_5$ torsion part survives to \Ktheory{} or vanishes
(there is no subgroup except the trivial group). But according to the
previous section there exists a torsion subgroup. Therefore
$K(X)_\mathrm{tors}=\Z_5$. Using Chern isomorphism and duality, this
determines \Ktheory{} completely:
\begin{equation}
  K^i(X) = 
  \left\{
    \begin{array}{cl}
      \Z^{44} \oplus \Z_5           & i=1 \\
      \Z^{4} \oplus \Z_5            & i=0 \\
    \end{array}
  \right.
\end{equation}

\subsection{The Tian--Yau manifold}

The first known example was the the Tian--Yau threefold $X = Y/\Z_3$,
a complete intersection $Y$ in $\CP{3}\times \CP{3}$ with a free
$\Z_3$ group action.  

The Hodge diamond can be found via counting monomials
(see~\cite{threegen} for details):
\begin{equation}
  h^{p,q}(Y) = \quad
  \begin{array}{ccccccc}
    & & &1& & & \\
    & &0& &0& & \\
    &0& &14& &0& \\
   1& &23& &23& &1\\
    &0& &14& &0& \\
    & &0& &0& & \\
    & & &1& & & 
  \end{array}
  \qquad
  h^{p,q}(X) = \quad
  \begin{array}{ccccccc}
    & & &1& & & \\
    & &0& &0& & \\
    &0& &6& &0& \\
   1& &9& &9& &1\\
    &0& &6& &0& \\
    & &0& &0& & \\
    & & &1& & & 
  \end{array}
\end{equation}
Since the fundamental group of the quotient $\pi_1(X)=\Z_3$ we also
know the torsion part $H_1(X,\Z)=\Z_3$. It remains to determine the
torsion part of $H^3(X,\Z)_\mathrm{tors} \simeq
H^4(X,\Z)_\mathrm{tors} \simeq H_2(X,\Z)_\mathrm{tors}$. 

But in this case the sequence eq.~(\ref{eq:eilenbergmclane}) does not
suffice. Again the group homology $H_2(\Z_3)=0$, but now it is unclear
what $\Sigma_2$ is. All we know is $\pi_2(X) = \pi_2(Y) = \Z^{14}$, and
the free part $H_2(X,\Z)_\mathrm{free} = \Z^6$. There is no reason for
the image $\pi_2(X)\to \Sigma_2$ not to contain a torsion part.

Therefore
\begin{equation}
  \label{tianyau}
  H^i(X,\Z) = 
  \left\{
    \begin{array}{cl}
      \Z               & i=6 \\
      \Z_3             & i=5 \\
      \Z^6 \oplus T    & i=4 \\
      \Z^{20} \oplus T & i=3 \\
      \Z^6\oplus \Z_3  & i=2 \\
      0                & i=1 \\
      \Z               & i=0 \\
    \end{array}
  \right.
\end{equation}
where $T$ is the unknown torsion part. The Atiyah--Hirzebruch spectral
sequence not only kills torsion subgroups in cohomology, it also puts
them together differently via extensions. So we know that \Ktheory{}
has torsion, but cannot determine the groups.

\section{Conclusion}

As we have explicitly seen the order of \Ktheory-torsion and
cohomology torsion is in general different. Thus substituting integer
cohomology for \Ktheory{} not only leads to the wrong charge addition
rules, it also does not yield the correct number of charges. Although
not being totally independent, one must consider the whole spectral
sequence connecting them.

This implies that discrete torsion on the field theory level must be
different from the \Ktheory{} interpretation of D-brane charges. The
most promising idea for a complete treatment is trying to find a
pairing (preferably a perfect pairing) between \Ktheory{} and
something else (maybe again \Ktheory) to $U(1)$ and use this to
construct a suitable partition function, as in~\cite{w2}\cite{wm}.

The whole discussion might be even relevant to the real world, since
phenomenologically interesting string compactifications need finite
non-zero $H_1(X,\Z)$ in order to further break the gauge group via
Wilson lines. But by the universal coefficient theorem,
$H^2(X,\Z)_\mathrm{tors} \simeq H_1(X,\Z)_\mathrm{tors}$, so torsion
charges appear in all realistic compactifications.


\begin{thebibliography}{}
\bibitem{dmw}
  D. Diaconescu, G. Moore, E. Witten,
  ``\textsl{$E_8$ Gauge Theory, and a 
    Derivation of \Ktheory{} from M-Theory}'', 
  hep-th/0005090
\bibitem{w1}
  E. Witten,
  ``\textsl{D-branes and \Ktheory}'', 
  JHEP \textbf{9812} (1998), hep-th/9810188
\bibitem{w2}
  E. Witten,
  ``\textsl{Duality Relations Among Topological Effects In String Theory}'',\\ 
  hep-th/9912086
\bibitem{abs}
  M. F. Atiyah, R. Bott, A. Shapiro,
  ``\textsl{Clifford Modules}'', 
  Topology \textbf{3} (1964) pp. 3
\bibitem{hk}
  A. Hanany, B. Kol,
  ``\textsl{On Orientifolds, Discrete Torsion, Branes and M Theory}'',\\
  hep-th/0003025
\bibitem{wm}
  E. Witten, G. Moore, 
  ``\textsl{Self-Duality, Ramond-Ramond Fields, and K-Theory}'',\\ 
  hep-th/9912279
\bibitem{k}
  M. F. Atiyah, 
  ``\textsl{\Ktheory}'', Benjamin (New York) 1967
  \\
  T. Friedrich, 
  ``\textsl{Vorlesungen \"uber K-Theorie{}}'',
  Teubner Verlag
\bibitem{sequence}
  M. F. Atiyah, F. Hirzebruch,
  ``\textsl{Vector Bundles and Homogenous Spaces}'',\\
  Proceedings of Symposia in Pure Mathematics \textbf{3} 1961 p. 7--38
\bibitem{kunneth}
  M. F. Atiyah, 
  ``\textsl{Vector Bundles and the K\"unneth Formula}'',\\
  Topology \textbf{1} 1962 p. 245--248
\bibitem{threegen}
  B. Greene, K. Kirklin, P. Miron, G. Ross,  
  ``\textsl{A Three-Generation Superstring Model (I)}'',\\
  Nucl. Phys. B \textbf{278} 1986 p. 667--693
\bibitem{rings}
  P. Aspinwall, D. Morrison,  
  ``\textsl{Chiral Rings Do Not Suffice: $N=(2,2)$ Theories with
    Nonzero Fundamental Group}'',\\
  hep-th/9406032, Phys. Lett. B \textbf{334} 1994 p. 79-86
\bibitem{eilenbergmclane}
  S. Eilenberg, S. MacLane,  
  ``\textsl{Relations Between Homology and Homotopy Groups of Spaces}'',\\
  Ann. Math. \textbf{46} 1945 p. 480--509  
\bibitem{bott_tu}
  R. Bott, L. Tu,  
  ``\textsl{Differential Forms in Algebraic Topology}'',\\
  Springer Verlag 1982
\end{thebibliography}
\end{document}